\documentclass[aps,showpacs,twocolumn,preprintnumbers,amsmath,amssymb]{revtex4}
\usepackage{graphicx}
\usepackage{bm}

\def\br{ \bm{r} }
\def\bk{ \bm{k} }

\def\bA{ \bm{A} }
\def\bB{ \bm{B} }

\begin{document}

\title{Phenomenological theory of superconductivity near domain walls in ferromagnets}

\author{K.~V.~Samokhin and D. Shirokoff$^*$}

\affiliation{Department of Physics, Brock University,
St.Catharines, Ontario, Canada L2S 3A1}
\date{\today}

\begin{abstract}
We develop a phenomenological model of superconductivity near a
domain wall in a ferromagnet. In addition to the electromagnetic
interaction of the order parameter with the ferromagnetic
magnetization, we take into account the possibility of a local
enhancement or suppression of superconducting pairing in the
vicinity of the wall, and also a non-perfect transparency of the
wall to electrons. It is found that the critical temperature of
superconductivity near the domain wall might be substantially
higher than in the bulk.
\end{abstract}


\maketitle

\section{Introduction}
\label{sec: Intro}

The recent discovery of superconductivity co-existing with
ferromagnetism in UGe$_2$ \cite{UGe2 exp}, ZrZn$_2$ \cite{ZrZn2
exp}, and URhGe \cite{URhGe exp} have revived the long-standing
interest to the interplay of the two phenomena. Although the
microscopic mechanism responsible for superconducting pairing in
the presence of a ferromagnetic background is still subject to
considerable controversy, some useful information about the
properties of these systems can be obtained from a
phenomenological approach, based on general symmetry arguments and
the Ginzburg-Landau (GL) equations, see e.g. Ref. \cite{Book}. A
symmetry-based description of ferromagnetic (FM) superconductors
has been developed in Refs.
\cite{MO01,Fom01,WS02,SW02-1,SW02-2,Sam02,Min02}.

In contrast to the non-magnetic case, a phenomenological theory of
superconductivity in the presence of ferromagnetism should include
the effects of internal magnetic induction $\bB$ created by the
macroscopic FM magnetization $\bm{M}$. That the coupling of the
orbital motion of electron charges with $\bB$ would suppress
superconductivity even in zero external field was pointed out by
Ginzburg \cite{Ginz56}, who argued that superconductivity could
not exist in any of the FM materials known at that time. While
valid in a single-domain ferromagnet, his argument should be taken
with care if $\bB$ is non-uniform. In order to minimize the
magnetic field energy, it is energetically favorable for a
macroscopic ferromagnet to break up into domains of size $L$ with
opposite directions of the magnetization $\bm{M}$, separated by
domain walls (DWs) of thickness $l$. Due to the spatial variation
of the internal field, it is expected that the suppression of
superconductivity in the vicinity of a DW is not as strong as in
the bulk, as was first hypothesized in Ref. \cite{MS60}. The
quantitative analysis shows \cite{BBP84} that the superconducting
critical temperature near a DW is indeed higher than the bulk
critical temperature deep inside the domain. The reason is that
the effective GL potential near the wall has the same double-well
shape as near a superconductor-insulator interface, leading to a
local enhancement of superconductivity similar to the surface
sheath effect \cite{SJdG63}. The same model was recently extended
in Ref. \cite{BM03} to calculate the critical temperature of the
superconducting DW in an external magnetic field.

In all previous treatments of the superconducting DW problem it
was assumed that the superconducting coupling constant is the same
throughout the system. However, it can be expected that, whatever
the microscopic pairing mechanism, the conditions for the
formation of the Cooper pairs near the DW and far from it are
different. The FM magnetization affects the electrons via the
orbital interaction of the electron charges with the induction
$\bB$, the exchange band splitting, and also the Zeeman coupling
of the electron spins with $\bB$. Because of the spatial variation
of both the exchange field and the internal induction, all three
mechanisms could change both the single-electron spectrum and the
electron-electron interactions in the vicinity of the DW, compared
to the bulk of the domains. In particular, the strength of the
Cooper pairing could be locally enhanced or suppressed near the
wall. Also, the DW creates a potential barrier for electrons
travelling from one domain to the other. The possibility of a
non-perfect transparency of the barrier should therefore be
included in the theory.

We assume that the thickness of the wall is small compared with
the length scales of superconductivity: $l\ll\xi_0\ll\lambda_L$,
where $\xi_0$ is the coherence length and $\lambda_L$ is the
London penetration depth. Then, in addition to the usual (i.e.
single-domain) terms, the general phenomenological GL functional
includes interface contributions localized near the DW, which
describe the variation of the pairing strength and a finite
transparency of the DW for electrons. These additional terms
affect the boundary conditions for the superconducting order
parameter at the wall \cite{DeGennes66}, and the problem then
becomes formally similar to that of the localized
superconductivity near a planar crystalline defect, which has been
extensively studied in the context of the twinning plane
superconductivity in high-$T_c$ cuprates and other superconductors
\cite{KhB87,And87,Sam94}.

The purpose of the present article is to study the phase diagram
of the localized DW superconductivity in a general
phenomenological model. The article is organized as follows: In
Sec. \ref{sec: Model}, we formulate our model and derive the GL
equations with appropriate boundary conditions. In Sec. \ref{sec:
Phase diagram}, we calculate the critical temperature of the DW
superconductivity as a function of an external field. The
evolution of the shape of the superconducting nucleus in the
non-linear regime below the critical temperature is studied in
Sec. \ref{sec: Nonlinear}.

\section{The model}
\label{sec: Model}

We assume that the Curie temperature $T_{FM}$ is higher that the
critical temperature $T_{SC}$ and treat the system in the
mean-field approximation, so the superconductivity appears in the
presence of a static FM background. According to Refs.
\cite{WS02,SW02-1,SW02-2,Sam02}, the order parameter in a FM
superconductor transforms according to one of the irreducible
co-representations of the magnetic symmetry group in the normal
state. Here we assume that the order parameter has one component
$\psi(\br)$, which is always the case in, e.g., ZrZn$_2$, assuming
that only one of the exchange-split bands is superconducting (in
principle, the pair scattering can induce order parameters of the
same symmetry on different sheets of the Fermi surface, leading to
a more complicated form of the free energy functional than the one
used below). The spin quantization axis is chosen parallel to
$\bm{M}(\br)$, so that the pairing in both domains occurs in, say,
spin-up channel. Thus the superconducting order parameters are
$\Delta_+(\bk,\br)=\Delta^{(+)}_{\uparrow\uparrow}(\bk,\br)=\psi_+(\br)\phi(\bk)$,
$\Delta_-(\bk,\br)=\Delta^{(-)}_{\uparrow\uparrow}(\bk,\br)=\psi_-(\br)\phi(\bk)$,
where $\phi(\bk)$ is the basis function of the chosen
one-dimensional co-representation.

The bulk critical temperature is the same on both sides of the
wall and equal to $T_{c0}$. We focus here on the case
$l\ll\xi_0\ll L$, where $\xi_0$ is the superconducting coherence
length, which allows us to consider a single wall separating two
domains with opposite directions of magnetization. The GL free
energy density can be represented as a sum of the bulk
contributions from the two domains, and the interface contribution
due to the presence of the DW: $F=F_++F_-+F_{DW}$. Choosing the DW
to coincide with the plane $x=0$, we have
\begin{equation}
\label{F bulk}
    F_\pm=a(T-T_{c0})|\psi_\pm|^2+K|\bm{D}\psi_\pm|^2+\frac{1}{2}\beta|\psi_\pm|^4,
\end{equation}
where $\psi_+=\psi(x>0)$ and $\psi_-=\psi(x<0)$ are the order
parameters in the right and left domains respectively,
$D_i=-i\nabla_i+(2\pi/\Phi_0)A_i(\br)$,
$\bm{\nabla}\times\bA=\bB$, $\Phi_0=\pi\hbar c/e$ is the magnetic
flux quantum (the electron charge is equal to $-e$). The internal
magnetic induction is given by $\bB=4\pi\bm{M}+\bm{H}$, where
$\bm{H}$ is an external field. Since the DW thickness is much
smaller that $\xi_0$, we can approximate the non-uniform
magnetization by a step-like function:
$\bm{M}(\br)=(0,0,M_0\,\mathrm{sign}\,x)$. If the external field
is directed along the $z$ axis, we have
$\bA(\br)=(0,B_0|x|+Hx,0)$, where $B_0=4\pi M_0$ (we assume
$M_0>0$). For simplicity, we neglect the anisotropy of the
effective mass tensor in the gradient terms.

The DW contribution, which is localized near $x=0$, can be written
in the following general form:
\begin{eqnarray}
\label{F wall}
    F_{DW}&=&[\gamma_1(|\psi_+|^2+|\psi_-|^2)-
    \gamma_2(\psi_+^*\psi_-+\psi_-^*\psi_+)\nonumber\\
    &&+i\gamma_3(\psi_+^*\psi_--\psi_-^*\psi_+)]\delta(x),
\end{eqnarray}
with real coefficients $\gamma_i$. We keep only the terms
quadratic in $\psi_\pm$, which are needed to obtain linear
boundary conditions at $x=0$, see below. The expression (\ref{F
wall}) is consistent with the symmetry of the system: $F_{DW}$ has
to be real and invariant with respect to time reversal
($\psi_\pm\to\psi_\pm^*$) accompanied by the interchange of the
two domains ($+\leftrightarrow -$). The parameters of the GL
functional depend on the domain magnetization $\bm{M}$. In
particular, in the limit $\bm{M}\to 0$ (no domain wall), all
$\gamma_i$ should vanish. While the first two terms in Eq. (\ref{F
wall}) are similar to those discussed in the context of the
localized superconductivity at planar crystalline defects
\cite{And87}, the last one is unique to the present problem
because it requires time-reversal symmetry breaking. It turns out
however that we can choose $\gamma_3=0$ without any loss of
generality, because the second and third terms in Eq. (\ref{F
wall}) can be combined together by rotating the order parameter
phases separately in each domain: $\psi_\pm\to e^{\mp
i\theta/2}\psi_\pm$, where $\theta=\arg(\gamma_3/\gamma_2)$. For
the same reason, the sign of $\gamma_2$ is not important, in
particular one can always choose it to be positive: if
$\gamma_2<0$, then the gauge transformation $\psi_+\to\psi_+$,
$\psi_-\to e^{i\pi}\psi_-$ changes $\gamma_2\to-\gamma_2$.

In our phenomenological model we allow for the order parameter to
be discontinuous across the domain wall, the rationale being that
the DW creates a potential barrier for electrons and therefore may
behave similar to an extended Josephson junction. The interface
terms (\ref{F wall}) then translate into the linear boundary
conditions for $\psi_\pm$, see e.g. Ref. \cite{DeGennes66}. The
magnitude of the coupling between the domains, described by the
$\gamma_2$ term, depends on the transmission coefficient of the DW
for electrons, which in turn is determined by the thickness and
the structure of the wall. In the limit of zero transparency,
$\gamma_2=0$ and the domains are completely decoupled. The
continuity of the order parameter is restored in the limit
$\gamma_1=\gamma_2\to+\infty$, when the interface terms can be
written as $\gamma_1|\psi_+-\psi_-|^2$, which imposes the
condition $\psi_+(x=0)=\psi_-(x=0)$ [and also
$\psi'_+(x=0)=\psi'_-(x=0)$, see below]. If
$\gamma_1=\gamma_2+\delta\gamma$, with $\gamma_{1,2}\to+\infty$
and a finite $\delta\gamma$, then the order parameter is
continuous, and $F_{DW}=\delta\gamma|\psi|^2\delta(x)$. Depending
on the sign of $\delta\gamma$, this term describes either a local
enhancement ($\delta\gamma<0$) or suppression ($\delta\gamma>0$)
of superconductivity near the DW.

Far from the wall, the magnetic induction is uniform and the field
dependence of the superconducting critical temperature is given by
the standard upper critical field expression
$T_{c,bulk}=T_{c0}-2\pi KB_0/\Phi_0a$ (one should keep in mind
that $T_{c0}$ itself can be a function of the magnetization). Thus
the superconductivity in the bulk is suppressed by the internal
induction $B_0$ even in the absence of the external field
\cite{Ginz56}. At $T<T_{c,bulk}$, both domains are in the mixed
state.

Let us now calculate $T_c$ in the vicinity of the wall. We assume
that the phase transition into the superconducting state is of
second order at all fields. To find the phase diagram in the
external field $\bm{H}$, we use the linearized GL equations
obtained from Eq. (\ref{F bulk}), with the boundary conditions
supplied by the interface terms (\ref{F wall}). The order
parameter has the form $\psi_{\pm}(\br)=e^{iqy}f_\pm(x)$, where
the functions $f_{\pm}(x)$ satisfy the equations
\begin{eqnarray}
\label{GL eqs general}
    -K\frac{d^2f_\pm}{dx^2}+K\left[\frac{2\pi}{\Phi_0}(H\pm
    B_0)x+q\right]^2f_\pm\nonumber\\
    +a(T-T_{c0})f_\pm=0,
\end{eqnarray}
supplemented by the boundary conditions at $x=0$:
\begin{equation}
\label{bcs gen} \left\{
\begin{array}{l}
    \displaystyle
    K\frac{df_+}{dx}=\gamma_1f_+-\gamma_2f_-,\\
    \\
    \displaystyle K\frac{df_-}{dx}=\gamma_2f_+-\gamma_1f_-.
\end{array}\right.
\end{equation}

To simplify the notations, it is convenient to use dimensionless
variables: $\tilde x=x/l_B$, $\tilde q=ql_B$, and
$\tilde\gamma_i=\gamma_il_B/K$, where $l_B=\sqrt{\Phi_0/2\pi
B_0}$. Omitting the tildas, the GL equations take the form
\begin{equation}
\label{GL eqs}
    -f_\pm''+[(1\pm h)x\pm q]^2f_\pm+(\tau-1)f_\pm=0,
\end{equation}
where $h=H/B_0\geq 0$ is the dimensionless external magnetic
field, and $\tau=1+a(T-T_{c0})\Phi_0/2\pi B_0K$ is the
dimensionless temperature. It is easy to see that the zero-field
bulk critical temperature $T_{c,bulk}$ corresponds to $\tau=0$,
while $T_{c0}$ corresponds to $\tau=1$. The boundary conditions
(\ref{bcs gen}) become
\begin{equation}
\label{bcs}
    f_\pm'\mp\gamma_1 f_\pm\pm\gamma_2f_\mp=0.
\end{equation}

As seen from Eqs. (\ref{GL eqs}), the effective GL potential
$V(x)$ is described by two parabolas matched at $x=0$. We have to
find the absolute minimum of the ground state energy $1-\tau$ in
this potential as a function of the parameter $q$. The solution of
Eqs. (\ref{GL eqs}) that does not diverge (in fact, vanishes) at
$|x|\to\infty$, has the form
\begin{eqnarray}
\label{GL solution}
    f_\pm(x)=C_\pm e^{-X_\pm^2/2}
    H_{\nu_\pm}(\pm X_\pm),
\end{eqnarray}
where $C_\pm$ are some constants, $H_\nu(z)$ is the Hermite
functions \cite{Hermite}, and
$$
    \nu_\pm=\frac{1}{2}\left(\frac{1-\tau}{|1\pm h|}-1\right),
    \ X_\pm=\sqrt{|1\pm h|}\left(x\pm\frac{q}{1\pm h}\right).
$$
These expressions are singular at $h=\pm 1$, which requires
special consideration, see below. Substituting $f_\pm(x)$ in the
boundary conditions (\ref{bcs}) and using the fact that
$H'_\nu(z)=2\nu H_{\nu-1}(z)$, we obtain the solutions of two
types. First, there always exists a trivial bulk solution located
deep inside the domain with lower induction, for which $q=+\infty$
or $-\infty$ and $1-\tau=|1\pm h|$, thus the critical temperature
is simply $\tau_{c,bulk}(h)=1-\min|1\pm h|=1-|1-h|$ (at $h\geq
0$). Second, we can also have a non-trivial solution localized
near the domain wall, which is sensitive to the values of
$\gamma_i$. The corresponding equation for $\tau(h,q)$ is
\begin{widetext}
\begin{equation}
\label{Tc eq}
\left[L_{\nu_+}\left(q\frac{\sqrt{|1+h|}}{1+h}\right)-\frac{\gamma_1}{\sqrt{|1+h|}}\right]
    \left[L_{\nu_-}\left(q\frac{\sqrt{|1-h|}}{1-h}\right)-\frac{\gamma_1}{\sqrt{|1-h|}}\right]
    =\frac{\gamma_2^2}{\sqrt{|1-h^2|}}.
\end{equation}
\end{widetext}
Here we introduced $L_\nu(z)=2\nu H_{\nu-1}(z)/H_\nu(z)-z$. The
critical temperature $\tau_c(h)$ is given by the absolute maximum
of $\tau(h,q)$ with respect to $q$ at given $h$. Clearly
$\tau_c(-h)=\tau_c(h)$ [this is consistent with Eq. (\ref{Tc eq})
being invariant under $h\to-h$], therefore we assume $h\geq 0$.

As mentioned above, the case $h=1$ requires special care, because
of the singularities in Eq. (\ref{Tc eq}). Physically, this
corresponds to a complete compensation of the internal induction
by the external field in the domain $x<0$. While $f_+(x)$ in this
case is still expressed in terms of the Hermite functions,
$f_-(x)$ is given by a simple exponential. The equation for
$\tau(h=1,q)$ is
\begin{equation}
    \left[\sqrt{2}L_{\nu}\left(\frac{q}{\sqrt{2}}\right)-\gamma_1\right]
    \left(\sqrt{\tau-1+q^2}+\gamma_1\right)+\gamma_2^2=0,
\end{equation}
where $\nu=-(\tau+1)/4$ [note that the critical temperature of the
DW superconductivity is never below the bulk critical temperature,
in particular $\tau_c(h=1)\geq 1$].

\section{Phase diagram}
\label{sec: Phase diagram}

\subsection{Transparent DW}
\label{sec: case 1}

Before discussing the solution of Eq. (\ref{Tc eq}) for arbitrary
values of $\gamma_1$ and $\gamma_2$, let us first look at some
important particular cases. If the domain wall is completely
transparent for the electrons, then
$\gamma_1=\gamma_2+\delta\gamma$, $\gamma_{1,2}\to+\infty$, and
$f_+=f_-$. In this limit, Eq. (\ref{Tc eq}) reduces to
\begin{eqnarray}
\label{transparent DW}
    &&\sqrt{|1+h|}L_{\nu_+}\left(q\frac{\sqrt{|1+h|}}{1+h}\right)\nonumber\\
    &&+\sqrt{|1-h|}L_{\nu_-}\left(q\frac{\sqrt{|1-h|}}{1-h}\right)=2\,\delta\gamma.
\end{eqnarray}
In the simplest case $\delta\gamma=0$, corresponding to the
absence of the local enhancement or suppression of
superconductivity, the results of Ref. \cite{BM03} are recovered
from Eq. (\ref{transparent DW}). At zero external field we have
$\nu_\pm=\nu=-\tau/2$, and Eq. (\ref{transparent DW}) becomes
$L_\nu(q)=0$, whose numerical solution shows that the ground state
energy $1-\tau=2\nu+1$ reaches its minimum at $\nu=\nu_*\simeq
-0.20$ and $q=q_*\simeq -0.77$. In terms of the critical
temperature, this gives $\tau_c(h=0)=-2\nu_*\simeq 0.41$ and
$T_c\simeq T_{c0}-0.59(2\pi KB_0/\Phi_0a)>T_{c,bulk}$. Thus,
superconductivity can exist in the vicinity of the DW at a
temperature higher than $T_{c,bulk}$. This effect has the same
origin as the surface critical field $H_{c3}$ \cite{BBP84}. At
$h\neq 0$, the critical temperature first grows as a function of
$h$, because the external field partially compensates the internal
induction in one of the domains, and then finally starts to
decrease at $h>1$, see Fig. 1.

At $h=0$ but $\delta\gamma\neq 0$, we have
\begin{equation}
\label{transparent DW local SC}
    L_{-\tau/2}(q)=\delta\gamma
\end{equation}
If the superconductivity is locally suppressed, i.e.
$\delta\gamma>0$, then the critical temperature is always lower
than at $\delta\gamma=0$. In contrast, if the superconductivity is
locally enhanced, i.e. $\delta\gamma<0$, then $\tau_c$ is always
higher than at $\delta\gamma=0$.

The results of numerical solution of Eq. (\ref{transparent DW})
are presented in Fig. 1. The bulk critical temperature
$\tau_{c,bulk}(h)$ serves as the lower boundary for $\tau_c(h)$ in
the limit of strong suppression of superconductivity near the DW.
In the opposite limit of large negative $\delta\gamma$ (i.e.
strong superconductivity enhancement), the influence of the
variation of the internal induction is less pronounced, so that
the phase diagram becomes similar to that of a superconducting
plane defect in a non-magnetic crystal \cite{KhB87}. The order
parameter is continuous and localized in the vicinity of the DW,
see Figs. 2 and 3.

Another notable feature of the phase diagram is that, in contrast
to the upper critical field in the bulk, $\tau_c(h)$ for the DW
superconductivity is quadratic in field at $h\to 0$:
$\tau_c(h)-\tau_c(h=0)=O(h^2)$. This happens because the
localization of the order parameter near the DW makes it possible
to treat $h$ as a small perturbation, which cannot be done in the
bulk case.

\subsection{Decoupled domains}
\label{sec: case 2}

The opposite case $\gamma_1=\gamma_2=0$ corresponds to completely
decoupled domains with the order parameters obeying $f'_\pm(0)=0$.
As follows from Eq. (\ref{Tc eq}), the critical temperature in
this case is determined by the equation
$L_{\nu_-}[q\sqrt{|1-h|}/(1-h)]=0$ at $h\geq 0$, which gives
$\tau_c(h)=1-(2\nu_*+1)|1-h|\simeq 1-0.59|1-h|$, so we have an
$H_{c3}$-type behavior at all fields.

\subsection{General case}
\label{sec: case 3}

Let us now find the critical temperature and the shape of the
superconducting nucleus at zero external field for arbitrary
$\gamma_1$ and $\gamma_2$. At $h=0$ Eq. (\ref{Tc eq}) yields two
equations for the critical temperature:
$L_{-\tau/2}(q)=\gamma_\pm$, where
$\gamma_\pm=\gamma_1\pm\gamma_2$. Recalling that $\gamma_2\geq 0$,
the highest critical temperature is achieved when the right-hand
side is $\gamma_-=\delta\gamma=\gamma_1-\gamma_2$. At
$\delta\gamma=0$, the transparent DW case is recovered, see Sec.
\ref{sec: case 1}.

At arbitrary $h$, $\gamma_1$, and $\gamma_2$, Eq. (\ref{Tc eq}) is
solved numerically. A representative phase diagram is shown in
Fig. 4. In contrast to the transparent case, the order parameter
is now discontinuous across the DW, but it still is localized in
its vicinity, see Figs. 5,6. The qualitative features of the phase
diagram are quite similar to those discussed in Sec. \ref{sec:
case 1}. In particular, the crossover from the bulk regime to the
strongly-localized superconducting DW regime is controlled by
$\delta\gamma=\gamma_1-\gamma_2$.

\section{Non-linear effects}
\label{sec: Nonlinear}

We have also studied the evolution of the shape of the
superconducting nucleus as a function of temperature at zero
external field, assuming a transparent DW with no local
enhancement or suppression of superconductivity.

At $\gamma_1=\gamma_2=+\infty$, the order parameter and its
derivative are continuous across the wall. Writing the order
parameter in the dimensionless form: $\psi(\br)=\psi_0 e^{iqy}
f(x)$, where $\psi_0=\sqrt{K/\beta l_B^2}$, the free energy
becomes
\begin{equation}
\label{F nonlinear}
    \frac{F}{F_0}=\left(\frac{df}{dx}\right)^2+(\tau-1)f^2
    +(|x|+q)^2f^2+\frac{1}{2}f^4
\end{equation}
($F_0=K^2/\beta l_B^4$), which yields the following non-linear GL
equation:
\begin{equation}
\label{GL eq nonlinear}
    -f''+(|x|+q)^2f+(\tau-1)f+f^3=0
\end{equation}
The solution of the linearized version of this equation in Sec.
\ref{sec: case 1} gave the critical temperature $\tau_c\simeq
0.41$, corresponding to $q_*\simeq-0.77$.

The non-linear equation (\ref{GL eq nonlinear}) is solved
numerically at $0<\tau<\tau_c$ by the ``shooting'' method
\cite{Pang97}: We start at some large value of $|x|=x_0\gg 1$ on
both sides of the DW and then integrate the equation towards a
matching point near the origin (although in the absence of an
external field, the solution is symmetric about the DW, the
matching point was not necessarily chosen to be at $x=0$). For
each $\tau$ and $q$, the starting value of $f(\pm x_0)=0$ was
chosen, while the starting slope of $f'(\pm x_0)$ was allowed to
vary to find the solution, which is smooth everywhere. The upper
and lower bounds on the initial slopes were estimated using the
fact that at large $|x|$ the non-linear term in Eq. (\ref{GL eq
nonlinear}) is negligible, and $f(x)$ decays approximately as a
Gaussian. Then the bisection method was used to converge on the
starting slope that guaranteed continuity of the wave function. At
the next step, for each wave function an evaluation of the GL free
energy (\ref{F nonlinear}) was made, and the value of $q$ was
chosen to minimize $F$. The final results were found to be
independent of the choice of the matching point.

Our results for the order parameter are presented in Fig. 6. The
height of the wave function, i.e. $f(x=0)$, grows approximately as
$\sqrt{\tau_c(h=0)-\tau}$. As a check of the validity of our
procedure, the temperature dependence of the parameter $q$ was
found to converge perfectly to the previously calculated value
$q_*$ as $\tau$ approaches $\tau_c(h=0)$.

\section{Conclusions}

We have studied the phase diagram of a superconducting domain wall
in a ferromagnet using a phenomenological model featuring
additional interface terms in the GL energy. These terms are
introduced in order to account for the possible variation of the
superconducting coupling constant near the DW, and also for a
finite transparency of the wall to electrons. The critical
temperature for the localized DW superconductivity is found to be
always higher than in the bulk. Depending on the relation between
the phenomenological parameters in the interface terms, one can
get an enhancement of the critical temperature above its value for
a transparent wall, and also a non-monotonous dependence of the
critical temperature on the external field.

This work was supported by a Discovery Grant from the Natural
Sciences and Engineering Research Council (NSERC) of Canada
(K.S.), and also by the NSERC USRA Award (D.S.).

\newpage

\begin{figure}
    \label{fig: transparent}
    \includegraphics[angle=90,width=8cm]{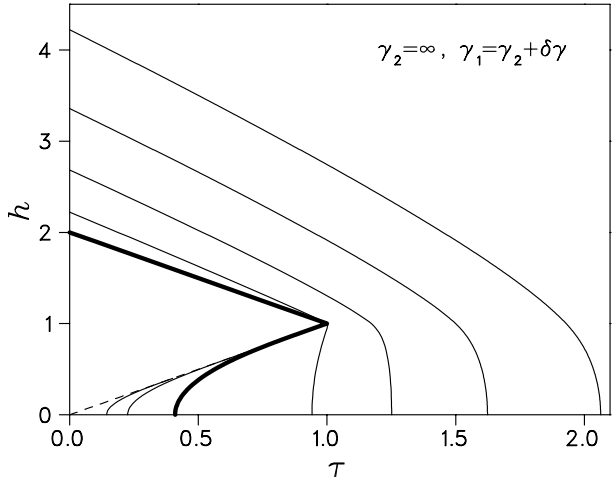}
    \caption{The superconducting phase transition lines for a transparent
    DW with a local enhancement or suppression of superconductivity, i.e.
    $\gamma_1=\gamma_2+\delta\gamma$, $\gamma_{1,2}\to+\infty$. The dashed line
    corresponds to the bulk critical temperature $\tau_{c,bulk}(h)=1-|1-h|$,
    the thick solid line -- to a transparent DW case with $\delta\gamma=0$,
    and the thin solid lines correspond to $\delta\gamma=0.5$, $0.3$,
    $-0.5$, $-0.7$, $-0.9$, $-1.1$ (from left to right).}
\end{figure}

\begin{figure}
    \label{fig: OP trans 1}
    \includegraphics[angle=90,width=8cm]{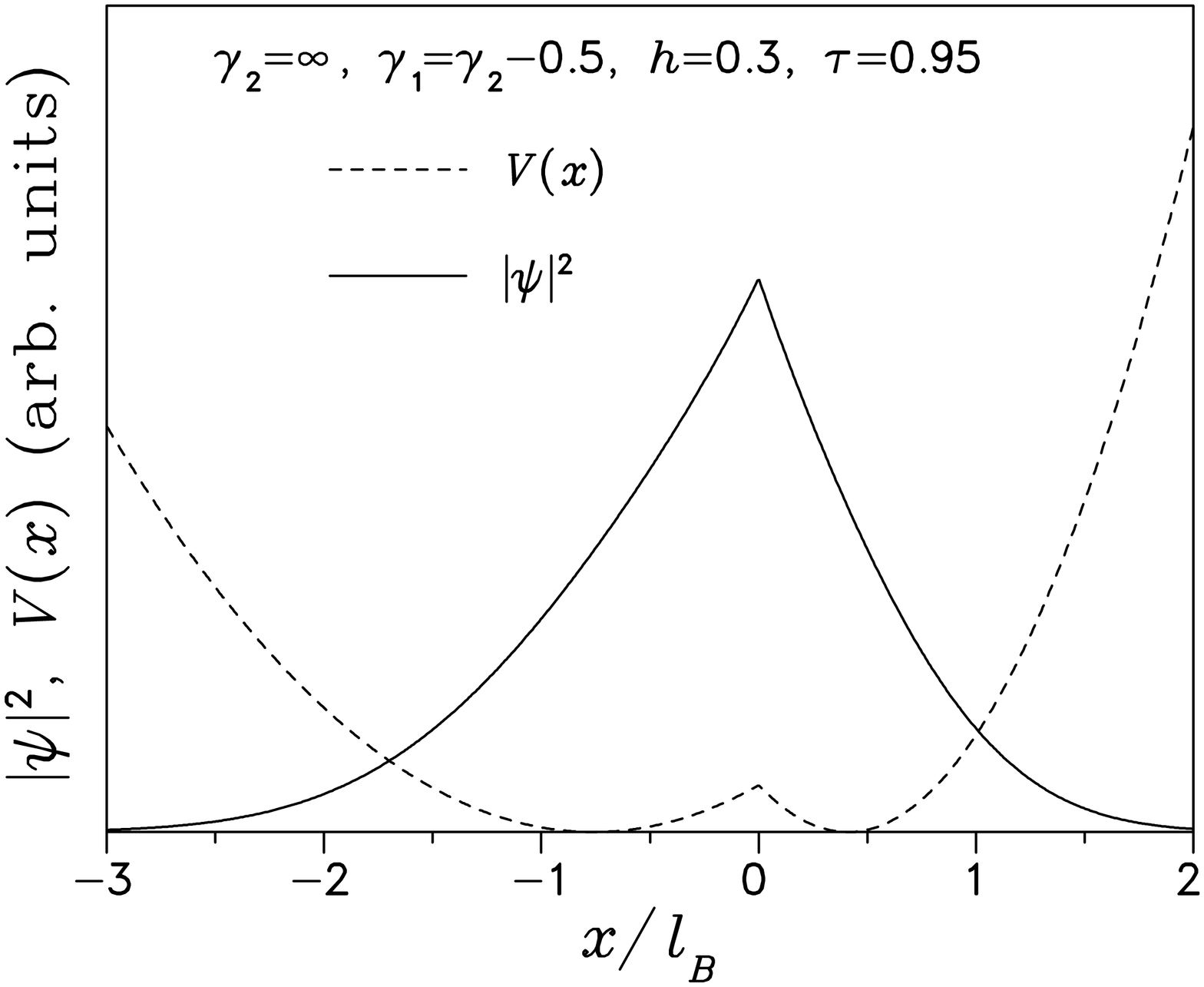}
    \caption{The superconducting order parameter and the GL
    effective potential $V(x)$ for $h=0.3$, at
    $\gamma_{2}=\infty$, $\delta\gamma=-0.5$.}
\end{figure}

\begin{figure}
    \label{fig: OP trans 2}
    \includegraphics[angle=90,width=8cm]{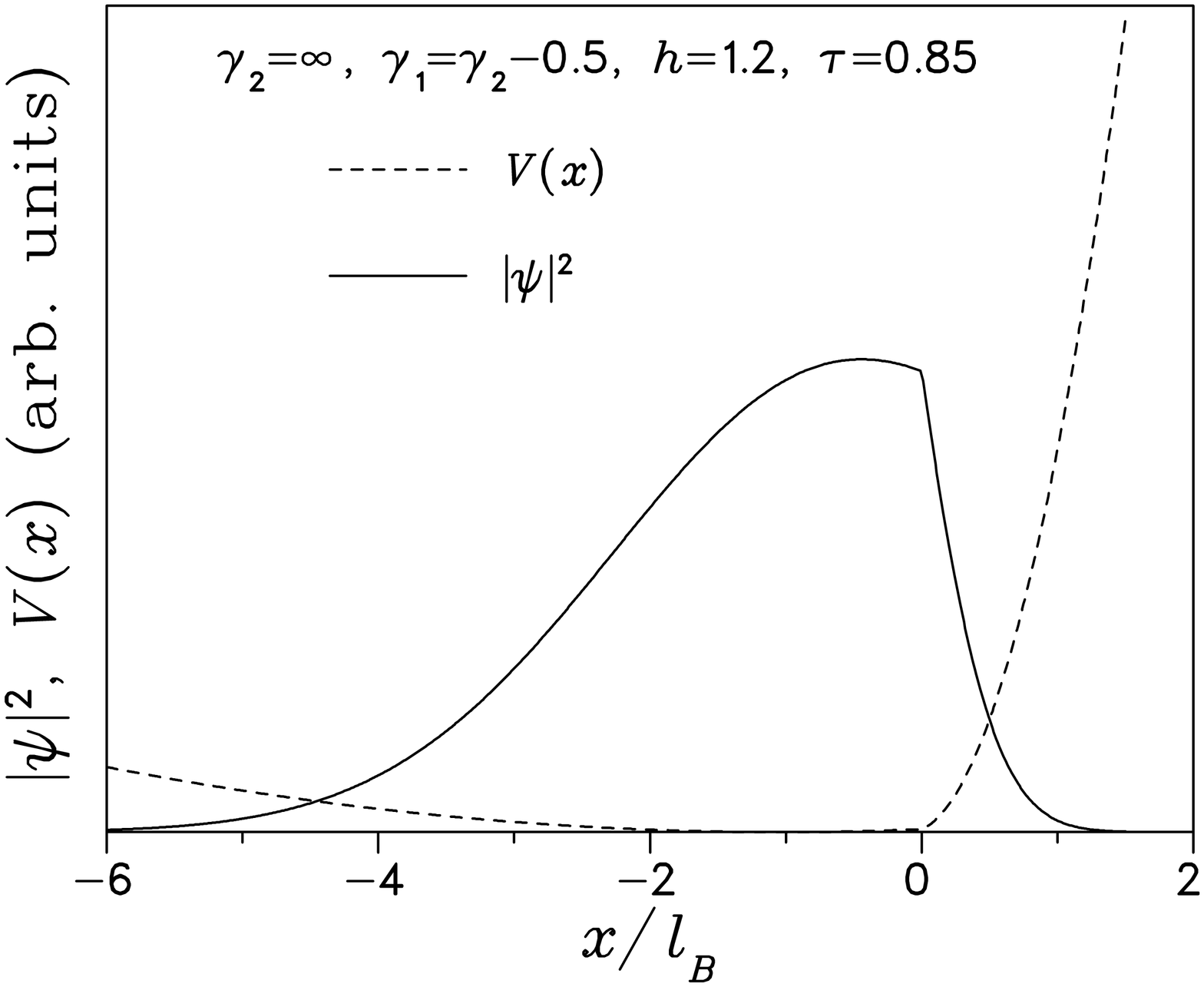}
    \caption{The superconducting order parameter and the GL
    effective potential $V(x)$ for $h=1.2$, at
    $\gamma_{2}=\infty$, $\delta\gamma=-0.5$.}
\end{figure}

\begin{figure}
    \label{fig: general}
    \includegraphics[angle=90,width=8cm]{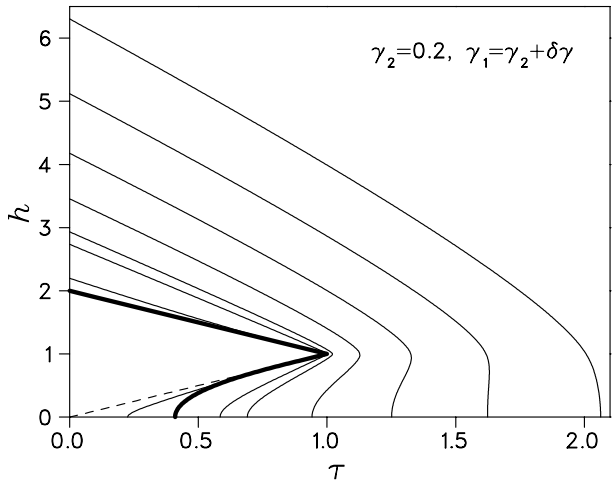}
    \caption{The superconducting phase transition lines for
    $\gamma_1=\gamma_2+\delta\gamma$, $\gamma_2=0.2$. The dashed line
    corresponds to the bulk critical temperature, the thick solid
    line -- to a transparent DW case with $\delta\gamma=0$, and the thin
    solid lines correspond to $\delta\gamma=0.5$, $0$,
    $-0.1$, $-0.3$, $-0.5$, $-0.7$, $-0.9$ (from
    left to right).}
\end{figure}

\begin{figure}
    \label{fig: OP gen 1}
    \includegraphics[angle=90,width=8cm]{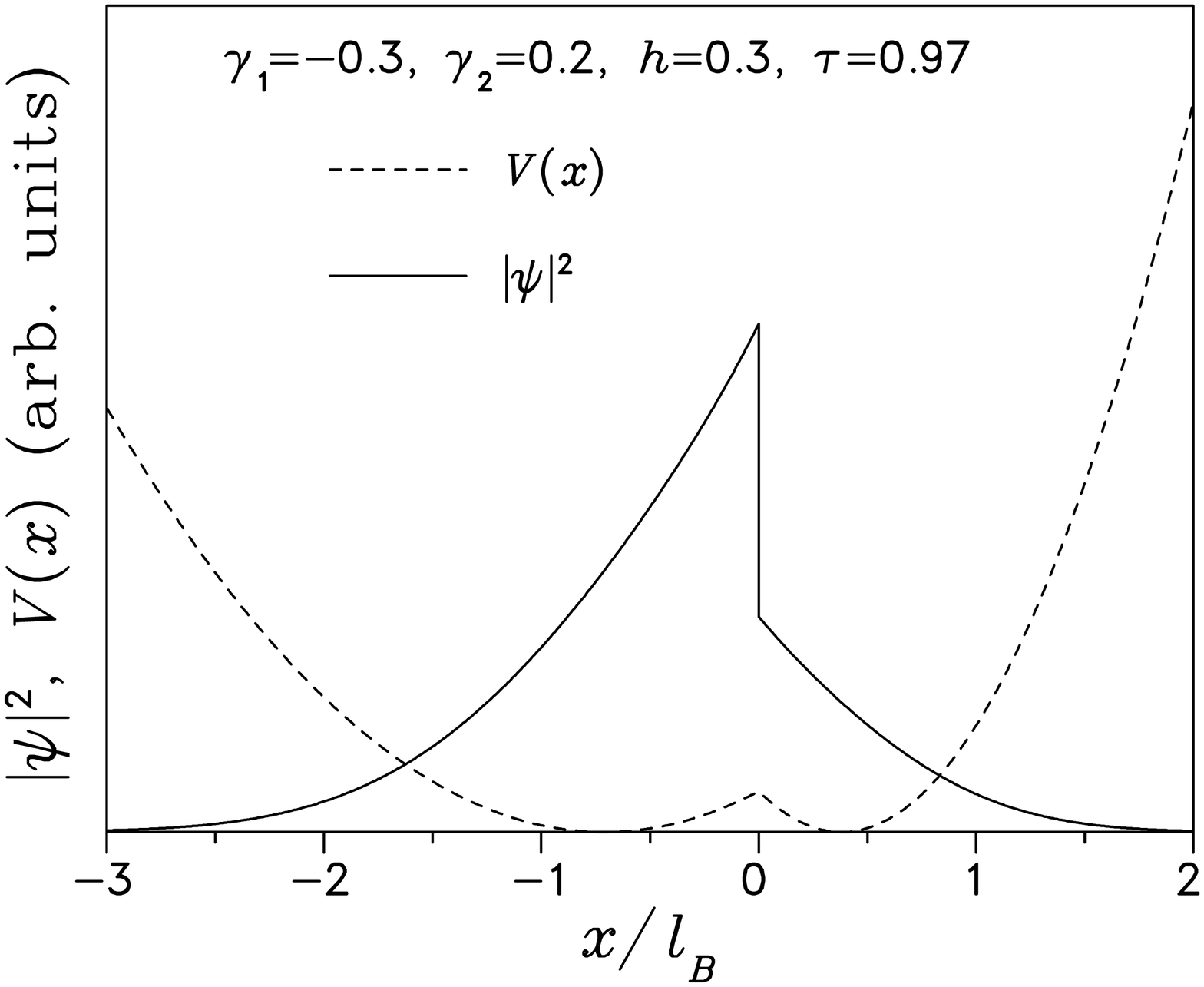}
    \caption{The superconducting order parameter and the GL
    effective potential $V(x)$ for $h=0.3$, at $\gamma_1=-0.3$,
    $\gamma_2=0.2$.}
\end{figure}

\begin{figure}
    \label{fig: OP gen 2}
    \includegraphics[angle=90,width=8cm]{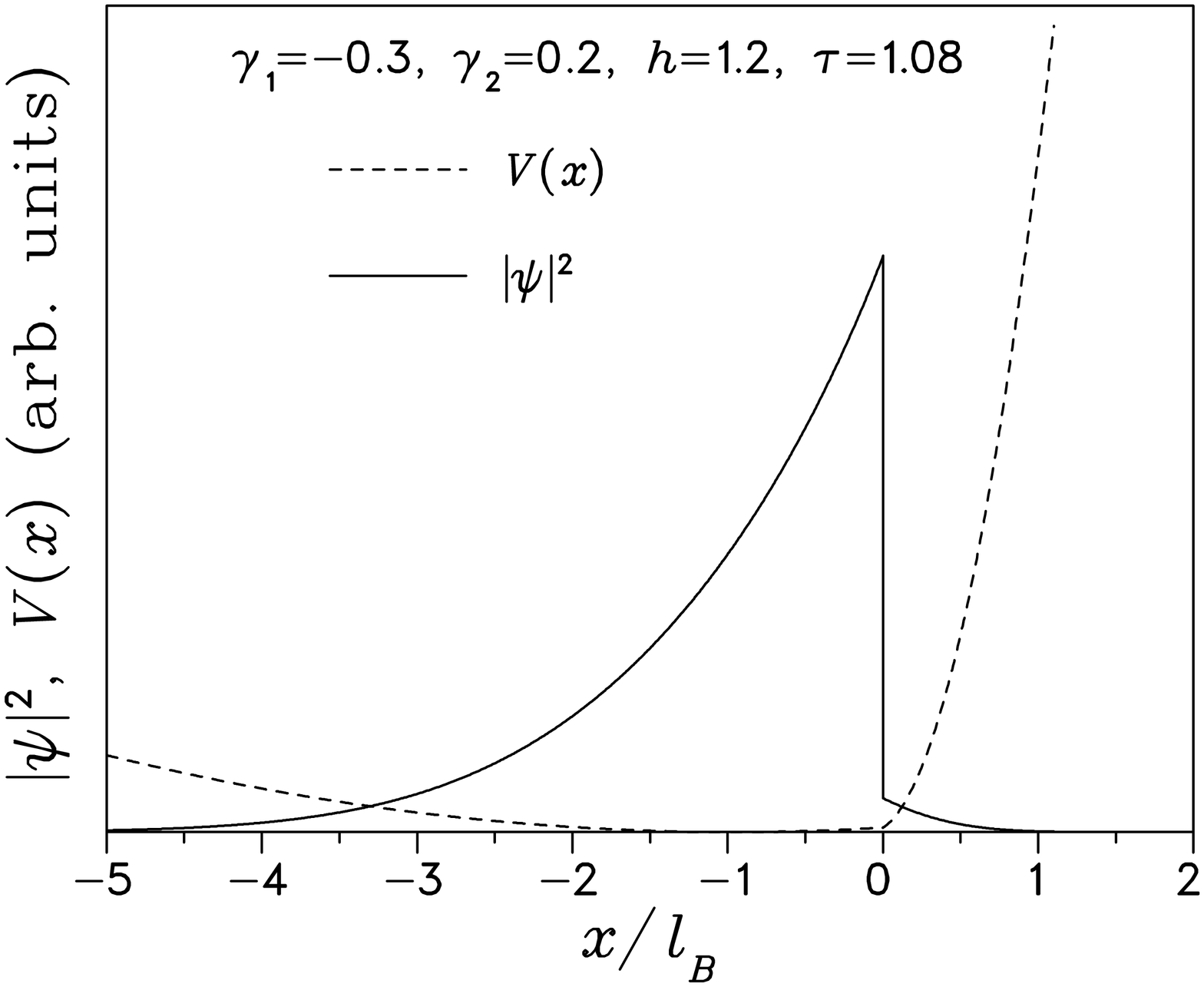}
    \caption{The superconducting order parameter and the GL
    effective potential $V(x)$ for $h=1.2$, at
    $\gamma_1=-0.3$, $\gamma_2=0.2$.}
\end{figure}

\begin{figure}
    \label{fig: OP nonlinear}
    \includegraphics[angle=90,width=8cm]{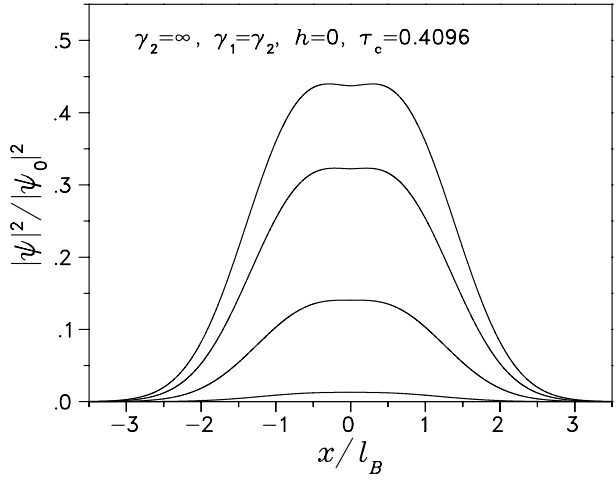}
    \caption{The superconducting order parameter in the non-linear regime below $\tau_c$, in the
    transparent DW case with no superconductivity enhancement [$\tau_c(h=0)\simeq 0.41$)].
    The curves correspond to $\tau=0.40$, $0.30$, $0.15$, and
    $0.05$ (from bottom to top).}
\end{figure}

\end{document}